\documentclass{article}
\usepackage{arxiv}

\usepackage[utf8]{inputenc} 
\usepackage[T1]{fontenc}    
\usepackage{xr}
\usepackage{xr-hyper}
\usepackage{hyperref}       
\usepackage{url}            
\usepackage{etoolbox}
\usepackage{multirow}
\usepackage{multicol}
\usepackage{booktabs}       
\usepackage{amsfonts}       
\usepackage{nicefrac}       
\usepackage{microtype}      
\usepackage{graphicx}
\usepackage{doi}
\usepackage[ruled,vlined]{algorithm2e}
\usepackage{amssymb,amsbsy,amsmath}
\usepackage{dcolumn}        
\usepackage[normalem]{ulem} 
\usepackage{nameref}
\usepackage[sort&compress]{natbib}
\usepackage{natmove}
\setcitestyle{super,square,citesep={,}, notesep={;}}
\usepackage{longtable}
\usepackage{zref-xr}
\zxrsetup{toltxlabel}
\usepackage{comment}
\usepackage{subfiles}

\DeclareMathOperator*{\argmax}{\arg\!\max}

\title{Iterative Symbolic Regression for Learning Transport Equations}

\author{ \href{https://orcid.org/0000-0001-5696-9193}{\includegraphics[scale=0.06]{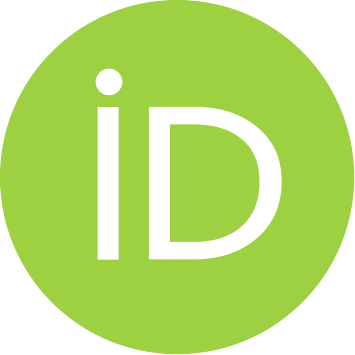}\hspace{1mm}Mehrad Ansari}\thanks{Denotes equal contribution} \\
	Department of Chemical Engineering\\
	University of Rochester\\
	Rochester, NY, 14627 \\
	\texttt{mgholiza@ur.rochester.edu} \\
	\And
	\href{https://orcid.org/0000-0002-9465-3840}{\includegraphics[scale=0.06]{orcid.pdf}\hspace{1mm}Heta A.~Gandhi$^*$}\\
	Department of Chemical Engineering\\
	University of Rochester\\
	Rochester, NY, 14627 \\
	\texttt{hgandhi@ur.rochester.edu} \\
	\And
	\href{https://orcid.org/0000-0000-0000-0000}{\includegraphics[scale=0.06]{orcid.pdf}\hspace{1mm}David G.~Foster} \\
	Department of Chemical Engineering\\
	University of Rochester\\
	Rochester, NY, 14627 \\
	\texttt{david.foster@rochester.edu} \\
	\And
	\href{https://orcid.org/0000-0002-6647-3965}{\includegraphics[scale=0.06]{orcid.pdf}\hspace{1mm}Andrew D.~White}\thanks{Corresponding Author} \\
	Department of Chemical Engineering\\
	University of Rochester\\
	Rochester, NY, 14627 \\
	\texttt{andrew.white@rochester.edu} \\
}

\date{}

\begin{document}

\maketitle
\begin{abstract}

Computational fluid dynamics (CFD) analysis is widely used in chemical engineering. Although CFD calculations are accurate, the computational cost associated with complex systems makes it difficult to obtain empirical equations between system variables. Here, we combine  active learning (AL) and symbolic regression (SR) to get a symbolic equation for system variables from CFD simulations. Gaussian process regression-based AL allows for automated selection of variables by selecting the most instructive points from the available range of possible parameters. The results from these experiments are then passed to SR to find empirical symbolic equations for CFD models. This approach is scalable and applicable for any desired number of CFD design parameters. To demonstrate the effectiveness, we use this method with two model systems. We recover an empirical equation for the pressure drop in a bent pipe and a new equation for predicting backflow in a heart valve under aortic insufficiency. 
\end{abstract}

\section*{Introduction}
Computational fluid dynamics (CFD) provides a numerical approximation to the conservation of mass, momentum and energy (Navier-Stokes equations) that govern the fluid flow behavior. While experimentally quantifying fundamental mechanisms of a process is often insightful, CFD modeling can provide an alternative approach for better understanding of the underlying physics in a less resource-intensive manner\cite{zawawi2018review}. With continued growth of computational power and advances in CFD techniques, even complex models can be done on commodity hardware. Thus, CFD modeling is routinely applied in several fields of science and engineering such as chemistry\cite{bornhorst2020urea, cai2015optimized}, materials\cite{subin2018analysis}, fluid flow and heat transfer\cite{mahian2019recent, tong2019review}, biology\cite{jayathilake2019modelling}, drug delivery\cite{koullapis2019multiscale}, semiconductors\cite{zhang2019multiscale}, environmental engineering \cite{lauriks2021application, collivignarelli2020identification}, biomedical engineering\cite{azriff2018haemodynamics} and aeronautics\cite{jun2018recent}.

Parametric analysis has been widely used in process modeling and design optimization using a semi-automated or fully-automated workflows\cite{bucklow2017automated, deininger2020continuous,gu2018automated}. However, running the entire CFD calculations on all possible design parameters can be computationally expensive, especially for complex CFD models. Thus, there is a need to identify which feature points are most important in experiment design, and allow for system analysis with fewer CFD simulations.  Another challenge in the mentioned settings is the lack of quantitative general equations that can be applied to different systems. This includes different geometrical designs, as well as having different operating conditions such as temperature, pressure, velocity or fluid properties. These system variables that can be inputs to CFD models are referred to as feature points in this work.

Here, we apply active learning (AL) to CFD modeling experiment design and then use symbolic regression (SR) to find empirical symbolic equations for these CFD models. AL is an iterative supervised learning technique that attempts to learn a good model from a few data points, by allowing the model to pick which data points it trains from.\cite{Settles2011} In other words, AL is the process of choosing the next experiment or feature point optimally using less resources and adding this new point to the training data.
In this paper, we use AL with gaussian process regression (GPR) to choose the next optimal CFD feature points. GPR is often associated with Bayesian optimization, where the goal is to optimize an expensive black box function. However, our goal is to find a symbolic equation across all feature values rather than a single optimum with as few CFD simulations as possible.

\begin{figure}
    \centering
    \includegraphics[width=\textwidth]{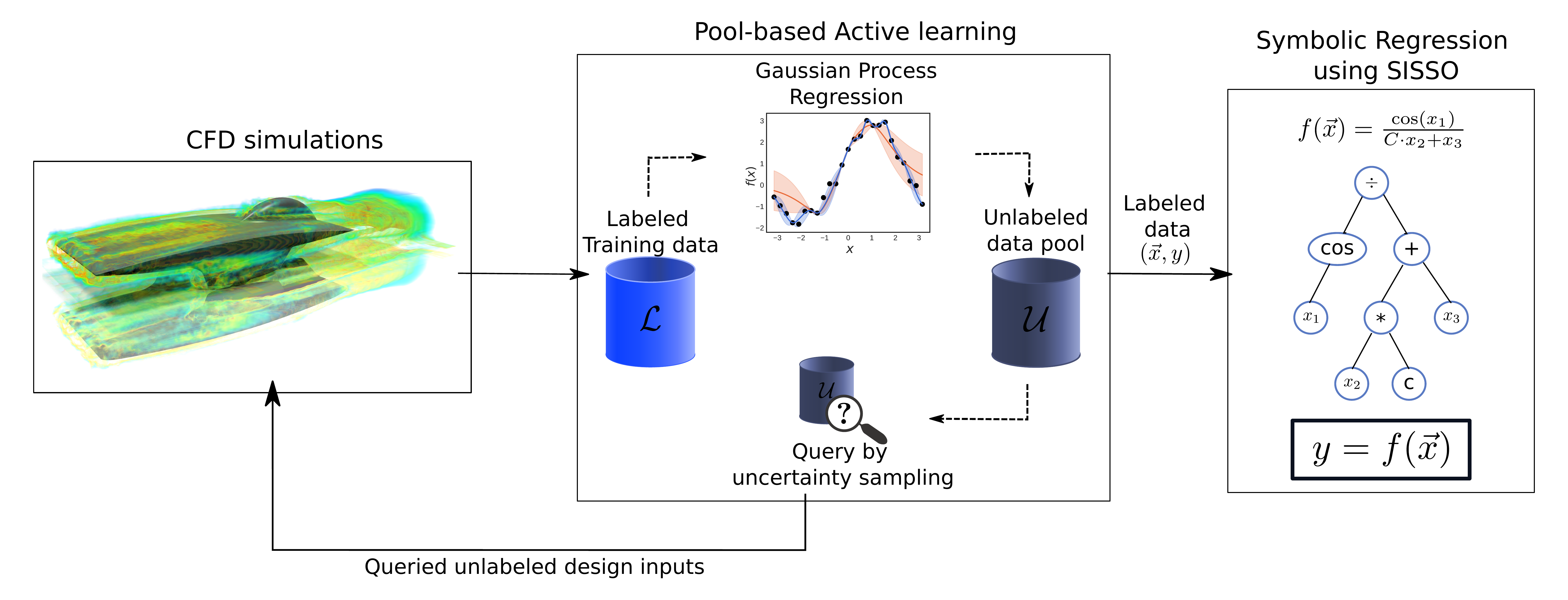}
    \caption{Process overview. A fully automated parameterized CFD model is coupled with pool-based active learning (AL). The CFD model is used to generate the labeled training data. The AL model is used to predict the next feature point for CFD simulations. An iteration of AL consists of learning the available data using gaussian process regression and then using uncertainty sampling to find the next feature point to label using a CFD simulation. The labeled data is then used as training data for symbolic regression to find the empirical symbolic equation for CFD feature inputs and outputs.} 
    \label{fig:concept}
\end{figure}

SR is a machine learning approach used to systematically determine symbolic equations that fit certain data with an unknown underlying function.\cite{Voss1998sr, Schmidt2009sr, Brunton2016sr} Unlike regression, where data is fit to a pre-defined function, SR attempts to find both the model and model parameters simultaneously. Neural networks (NN) are a popular choice for learning from data. However, even though they can approximate any function, the output from neural networks is difficult to interpret and cannot be converted to an analytical function. SR gives interpretable symbolic equations from data, which makes this approach appealing. Here, ``interpretable'' means that the exact relationship between input features and outputs is known in equation form. There are limited studies that explore SR to find general relationships from CFD simulations.\cite{Neumann2020, Chakraborty2020} In this study, we demonstrate the use of AL for design of CFD experiments, and then apply the SISSO SR method to determine the physics of fluid systems. Figure~\ref{fig:concept} provides an overview of this method. A fully-automated workflow is combined with AL to generate CFD data, which is then used to get an empirical symbolic equation using SR. To avoid non-physical symbolic equations, we include known asymptotic points using prior understanding of physics of the fluid systems being studied. These asymptotic points are included in the SR training data. This forces SR to return equations have the correct asymptotic behavior at extreme geometries and velocities. The AL and SR methods are described in detail in sections that follow.

\subsection*{Comparison to Related Work}
 Previous studies have implemented AL to accelerate simulation-driven design optimization. \citet{Owoyele2021} used AL to perform simulation-based data generation, ML learning and surrogate optimization to refine solution in the vicinity of predicted optimum parameters for design of a compression ignition engine. \citet{Goncalves2020} studied the generation of simulation-based surrogate models with the task of parameter domain exploration using various sampling and regression-based AL strategies. In a similar work, \citet{Pan2019}, used AL for developing surrogate models for industrial fluid flow case studies under a constraint of a limited function evaluations. AL has also been implemented in specific experiment design to deploy efficient design space  exploration to enhance model quality.\cite{Deng2018, vandermause2020fly}. Over the past few decades, multiple methods to solve the SR problem have been developed. Traditional deterministic algorithms assume a predefined mathematical function and attempt to find parameters with the best fit to the data, whereas, evolutionary algorithms try to find parameters and learn the best-fit function, simultaneously. Some prevalent methods are genetic programming algorithms \cite{Koza1994sr,Icke2013geneticprogramming, Lu2016geneticprogramming,Wang2019geneticprogramming, el2019solving, weatheritt2019improved, androulakis1991genetic}, sparse regression \cite{Brunton2016sr, Rudy2017sparseregression, schmelzer2020discovery, Naik2021sparseregression}, pareto-optimal regression \cite{Udrescu2020feynman, Udrescu2020feynman2}, and the sure-independence screening and sparsifying operator (SISSO) method.\cite{Ouyang2018sisso, Ouyang2019sisso} Most SR frameworks implement the popular Genetic programming, \cite{koza1992genetic} which is an improved version of Genetic Algorithms (GA), \cite{mitchell1998introduction,holland1992adaptation} inspired by Darwin's theory of natural selection. Genetic programming has also been used to identify hidden physical laws from the input-output response prior.\cite{vaddireddy2020feature, schmidt2009distilling, bongard2007automated} We use SISSO because it has been shown to be robust with small amounts of data.\cite{Ouyang2018sisso, xie2020machine, DeBreuck2021} This is advantageous for analysis of CFD systems where the computational cost of simulations increases with increasing number of variables and complexity of the system. Compared to existing work, our approach is novel in the sense of combining AL and SR to optimize training efficiency and output a general equation for any fluid system of interest.

\section*{Theory}

\subsection*{Governing Equations}
The governing equations are the conservation of mass, momentum and energy. With the assumption of steady state, incompressible flow and constant temperature, we have the continuity equation:
\begin{equation}
\label{eq:continuity}
    \nabla.\vec{v} = 0
\end{equation}
and the momentum equation can be simplified to:
\begin{equation}
\label{eq:momentum}
    \rho(\vec{v}.\nabla)\vec{v} = -\nabla P + \mu\nabla^2\vec{v}
\end{equation}
where, $\vec{v}$ is the velocity vector, $P$ is pressure and $\rho$ and $\mu$ denote the fluid density and viscosity, respectively. A Dirichlet boundary condition is imposed at the inlet, with a parabolic velocity profile normal to the boundary for all cases. This assumption allows the analysis for a fully-developed flow without the need for unnecessary geometry extensions, which results in extra mesh elements. The no-slip boundary condition imposed at the wall ensures a zero velocity relative to the pipe surface. Given the unknown pressure at the outlet, the outflow boundary condition is imposed. A convergence criterion is defined based on the conservation of mass at the inlet and outlet boundaries.

\subsection*{Pressure Drop in a Bent Pipe}
The laminar fluid flow in circular pipes is a classical problem in fluid mechanics, and it has been analyzed by the means of momentum balance, resulting in the famous Hagen-Poiseuille (HP) equation \cite{welty2020fundamentals}:
\begin{equation}
\label{eq:HP}
    w= \frac{\pi(P_0 - P)d^4\rho}{32\mu L}
\end{equation}
where the mass flow rate $w$ is the product of cross-sectional area, density and average velocity $\langle v\rangle$. Here $d$ and $L$ are the pipe diameter and length, respectively, and $P_0$ and $P$ denote the pressure at the inlet and outlet of the pipe. Note that Equation \ref{eq:HP} is only valid for continuous, laminar, incompressible, steady, Newtonian flow that is fully-developed. Given the pipe dimensions, fluid properties and average inlet velocity, one can easily obtain the pressure drop using Equation \ref{eq:HP}. Our goal here is to find an empirical equation for the pressure drop in a bent circular pipe as a function of the average inlet velocity ($\langle v \rangle$), pipe diameter ($d$) and  bend angle ($\theta$). The fluid is considered to be water at 25 $^{\circ}$C with constant properties. This setting is implemented to limit the number of feature points to three. However, more complicated models involving chemical reactions and convective heat transfer can be analyzed with more feature points. The geometry has been parameterized and meshed with hexahedral elements, as represented in Figure \ref{fig:bent_pipe}. More details on model parameterization can be found in Section~\ref{section1.1} of Supplementary Information (SI).

\begin{figure}[!ht]
    \centering
    \includegraphics[width=0.8\textwidth]{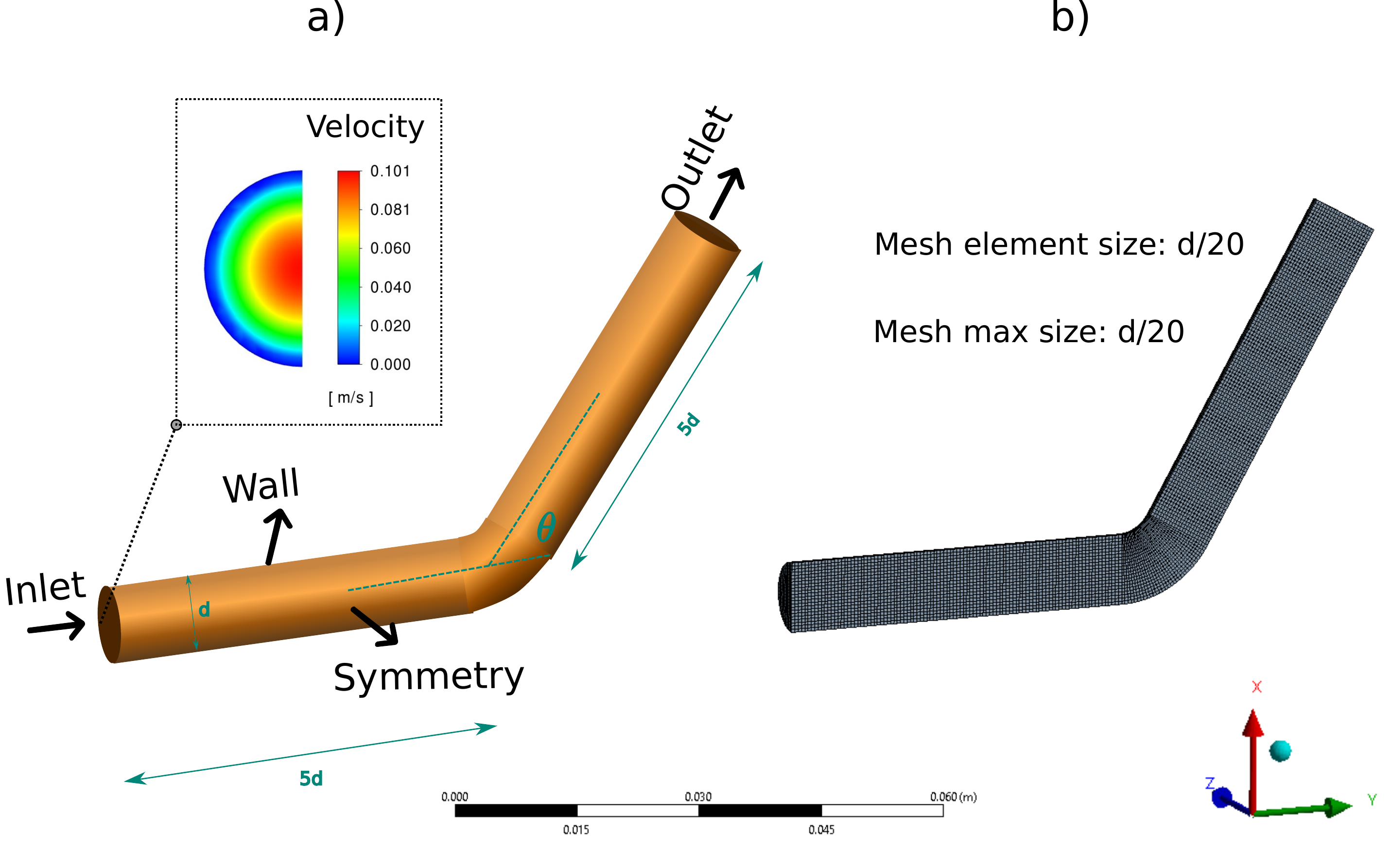}
    \caption{Bent pipe: a) parameterized geometry with inlet, wall, symmetry and outflow boundary conditions. A parabolic velocity profile is defined normal to the boundary to satisfy the assumption of fully-developed flow. The geometric constraints allow having different user inputs for $d$ and $\theta$ and ensure valid geometries. b) parameterized hexahedral mesh with element size and max element size of $d$/20, which allows for adjustable meshing given different geometric inputs for $d$ and $\theta$.}
    \label{fig:bent_pipe}
\end{figure}
Using Equation \ref{eq:HP}, the model is validated based on 5 different inputs with a bend angle of 1$^{\circ}$, which is approximately equivalent to a straight pipe. The mean error for the unit length pressure drop with respect to the HP equation is about 2$\%$.

\subsection*{Backflow at an Expansion Joint}

\begin{figure}[!ht]
    \centering
    \includegraphics[width=0.8\textwidth]{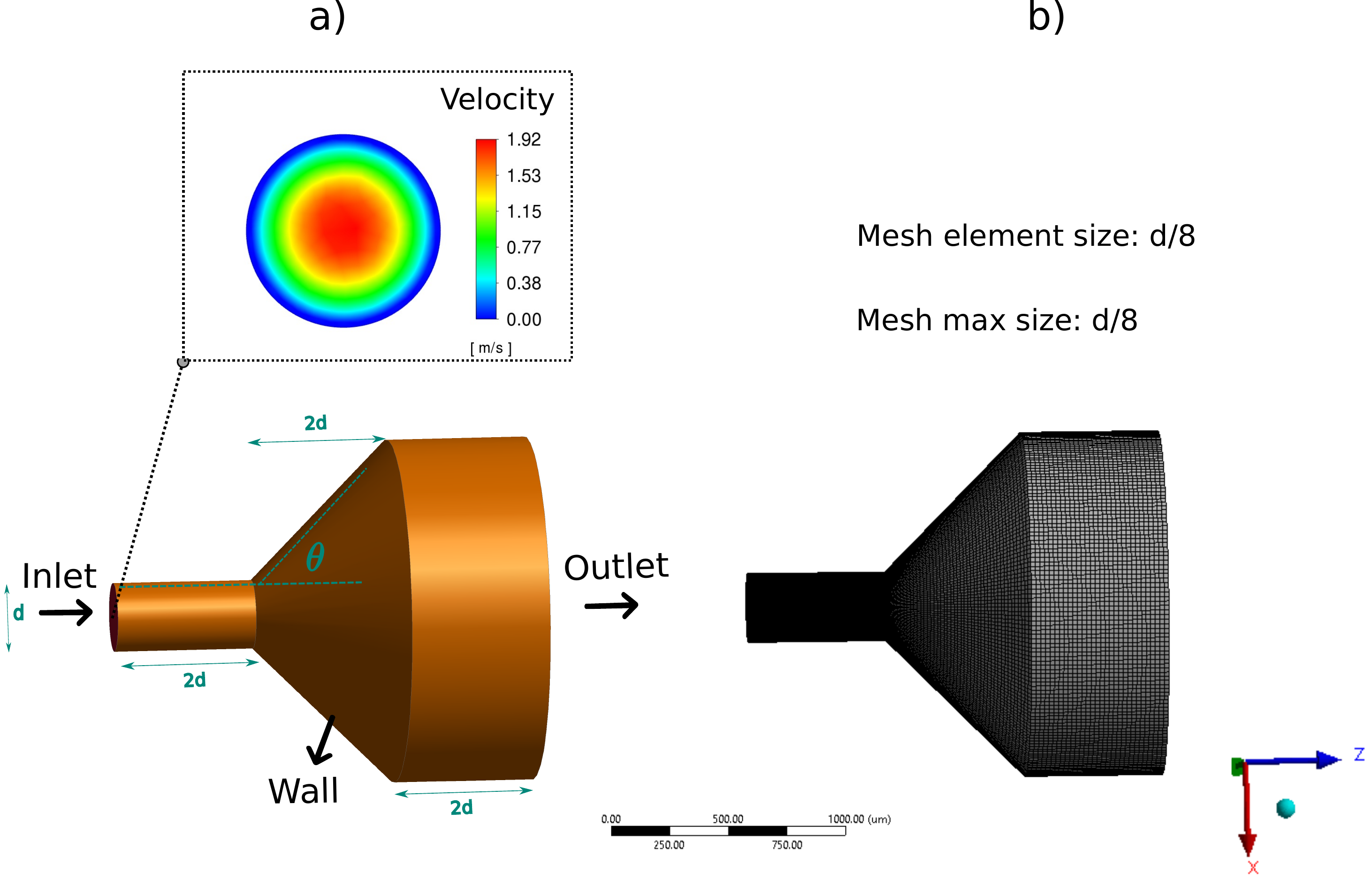}
    \caption{Expansion joint: a) parameterized geometry with inlet, wall and outflow boundary conditions. A parabolic velocity profile is defined normal to the boundary to satisfy the assumption of fully-developed flow. The geometric constraints allow having different user inputs for $d$ $\theta$ and ensure valid geometries.  b) parameterized hexahedral mesh with element size and max element size of $d$/8, which allows for adjustable meshing given different geometric inputs for $d$ and $\theta$.} 
    \label{fig:expansion_joint}
\end{figure}
We can consider the human heart to operate as two pumps in series. The right heart pumps blood to the pulmonic circulation and the left heart to the systemic circulation\cite{chandran2010role}. The valves in the human heart open and close efficiently, allowing the blood flow in the forward direction and minimizing the regurgitation of flood to the chamber it came from. Aortic insufficiency is a condition in which the heart valve fails to tightly close, allowing blood to flow backwards into the heart instead of pumping out \cite{prodromo2012aortic}. We have considered an expansion joint to simulate this condition in a simplified geometry and quantify the backflow volume. The fluid is blood with constant properties with density and viscosity set to 1060 Kg/m$^3$ and 0.004 Pa s at 37$^o$C. A fully-developed flow is defined at the inlet boundary, no-slip velocity at wall and outflow boundary condition at the outlet. The inlet pressure is set to 120 mmHg absolute. Once again, the geometry and the hexahedral mesh are constrained to avoid invalid models given different inputs (see Figure \ref{fig:expansion_joint}).  Blood dominantly flows in the $z$ direction, thus the $z$ component of the velocity is used as our metric for defining the backflow. The backflow volume is calculated by summing over the volume of mesh with negative velocity in $z$ direction (see SI Section~\ref{section1.2}).  The inputs to the model in this setting are the average inlet velocity ($\langle v \rangle$), inlet diameter ($d$) and expansion angle ($\theta$). The model outputs the percentage backflow ($f$) by finding the ratio of $V_{bf}$ to the total system volume.

\subsection*{Active Learning Model}
\label{AL}
The goal of AL algorithms is to increase accuracy of a machine learning model, while minimizing the training data required to train the model. It is often formulated as an optimization problem.\cite{Settles2011} Here, we use a pool-based AL setting. Pool-based AL assumes that the model has access to a large set of unlabeled samples. Consider a dataset $\mathcal{D} \supseteq \{\mathcal{L}, \mathcal{U}\}$ comprised of a small set of labeled data $\mathcal{L} = \{(\vec{x}_i, y_i)\}_{i=1}^{n_{\mathcal{L}}}$ with features $\vec{x}_i$ (n-dimensional vector) and corresponding labels $y_i$, and a large pool of unlabeled data $\mathcal{U} = \{\vec{x}_j\}_{j=1}^{n_{\mathcal{U}}}$ containing only features $\vec{x}_j$. $n_{\mathcal{L}}$ and $n_{\mathcal{U}}$ are the number of samples in the labeled and unlabeled dataset, respectively, and $n_{\mathcal{L}} \ll n_{\mathcal{U}}$. A model $\Phi$ is initially trained using the labeled data $\mathcal{L}$. Next, an unlabeled data sample, called a query, $\vec{x}_{i+1}$ is selected from the unlabeled data pool $\mathcal{U}$ using a query strategy. The selected query is then labeled by using an ``oracle''. An oracle is a human expert, or an experiment, but in this work it is a CFD simulation. The label $y_{i+1}$ is found by conducting a CFD simulation for the flow conditions and geometry specified by $\vec{x}_{i+1}$. This new observation $(\vec{x}_{i+1}, y_{i+1})$ is added to labeled data $\mathcal{L}$ and the model $\Phi$ is retrained based on this updated value. The process of query-label-retrain is iteratively repeated until the labeling budget is exhausted. 

In this study, we use gaussian process regression\cite{Rasmussen2006} (GPR) as our predictive model $\Phi$ and uncertainty sampling\cite{Lewis1994} as the query strategy. The uncertainty sampling strategy selects the feature point whose prediction is most uncertain. The model has the least information in the vicinity of the most uncertain prediction, and it is relatively more confident about other predictions. Hence, labeling the most uncertain point is most informative for the model.\cite{Settles2011, Nguyen2021} GPR is a probabilistic estimation method and has the ability to provide uncertainty measurement because it provides confidence intervals for predictions at each feature point. The goal of any regression model is to fit a function to datapoints. There are infinitely many functions that can possible fit a set of points. GPR assigns a probability to each of these functions.\cite{Kapoor2007} Once the GPR model  $\Phi$ is trained, the uncertainty in predictions and next feature point choice are calculated using Equations~\ref{eq:uncertainty} and \ref{eq:next_point}:
\begin{gather}
    U(\vec{x}) = \left[ 1 - \max\left(P^i_{\Phi}(y|\vec{x})\right) \right] \label{eq:uncertainty} \\
    \vec{x}_{i+1} = \argmax_x U(x). \label{eq:next_point}
\end{gather}
To ensure the method is robust to label noise, $U(x)$ is itself made a probability and $x_{i+1}$ is sampled rather than computing the $\argmax$. This is accomplished by computing the softmax of the predicted uncertainty, $U(x)$.\cite{Goodfellow2016} Thus, our AL equation is
\begin{equation} 
    P(\vec{x}_{i+1}) = \mathrm{softmax} [U(\vec{x})] =  \frac{e^{U(x)_j}}{\Sigma_{j=1}^{n_{\mathcal{U}}} U(x)_j} \label{eq:uncertainty_sampling} 
\end{equation}

Readers are referred to Section~\ref{sectionAL} of the SI for further details.

\subsection*{Symbolic Regression Model}
\label{SR}
To learn an interpretable model from CFD simulations, we use SISSO, developed by \citet{Ouyang2018sisso} SISSO aims to construct a symbolic equation between primary features $\vec{x}$ and labels $y$. Given M samples, SISSO assumes that the labels can be expressed as a linear combination of non-linear functions of primary features. So, $y = f(\Psi)$ where $\Psi = \{\psi_1, \psi_2, ...,\psi_r\}$ is a set of secondary features. The secondary features $\psi_i$ are non-linear, closed form functions of primary features. If $\vec{x} = [x_1, x_2, x_3, ..., x_n]$ are primary features, then examples of secondary features are $\{x_1/x_3,\  x_3 - x_1 x_2,\  x_4 x_5/x_1, ...\} $. 
These secondary features are obtained by recursively applying a set of user-defined operators on the primary features and creating a set of potential secondary features. The operator set can be any combination of unary and binary operators. The number of potential secondary features is proportional to the number of primary features used, the number of operators used, and the level of recursion. At each iteration, SISSO selects the subsets of secondary features that have the largest linear correlations with $y$. The number of terms in the linear expansion $f(\Psi)$ (called descriptors) are controlled by a sparsifying $l_0$ regularization. Note, here the number of descriptors refers to the number of terms in the output equation. For each iteration $q$, SISSO constructs multiple models for $f(\Psi)$ using the secondary features and selects the one with the largest correlation with the target property. 
More details on this procedure can be found in \citet{Ouyang2018sisso} 

In SISSO, dimensional analysis is performed to retain only valid combinations of primary features. This ensures that secondary features do not have unphysical units (e.g. \textit{force + time}). To achieve this, there is an option in SISSO to group primary features that have the same derived units. 
We modified this option so that the primary features are expressed in terms of fundamental units of measurement (mass, length, time, angle) and grouped based on these fundamental units instead of derived units. 


\section*{Methods}

In this section, the fully-automated CFD workflow, AL and symbolic regression procedures are explained. As seen in Figure~\ref{fig:concept}, we first use AL to get labeled data from the CFD model and then perform SR using that as training data, to obtain an empirical symbolic equation between features and labels. 

Our fully-automated workflow has been used on two different fluid flow problems, as described in previous sections. These problems are not necessarily complex, and the main goal here is to demonstrate a more robust approach that can also be applied to complex problems. In this work, we have coupled ANSYS Workbench with python. The parameterized CFD models are developed by defining input feature points. These inputs can include geometric features, operating conditions and fluid flow properties. They are easily adjustable in a python script and the outputs are updated accordingly. 

\newcommand{\colwidth}{0.16}
\newcommand{\smallcol}{0.15}
\begin{table}
    \caption{Feature ranges and test data split for the two systems.}
    \centering
    \resizebox{\textwidth}{!}{
    \begin{tabular}{m{0.2\textwidth}*{4}{m{\smallcol\textwidth}}*{2}{m{\colwidth\textwidth}}}
        \toprule
         \multirow{2}*{\parbox{\colwidth\textwidth}{\textbf{System}}} & & \multirow{2}*{\parbox{\colwidth\textwidth}{\textbf{Features}}} &  &
         \multirow{2}*{\parbox{\smallcol\textwidth}{\textbf{Labels}}} &
         \multirow{2}*{\parbox{\colwidth\textwidth}{\textbf{Test data}}} &
         \multirow{2}*{\parbox{\colwidth\textwidth}{\textbf{Data Split (train:test counts)}}} \\\\
         \cmidrule(l){2-4}
         & \multicolumn{1}{>{\centering\arraybackslash}m{\smallcol\textwidth}}{$d$ (m)} & \multicolumn{1}{>{\centering\arraybackslash}m{\smallcol\textwidth}}{$\theta\  (^\circ)$} & \multicolumn{1}{>{\centering\arraybackslash}m{\smallcol\textwidth}}{$\langle v_{in} \rangle$ (m/s)}\\
         \midrule
        Bent pipe  & $0.005 - 0.1$ & $1 - 180$ & $0.005 - 0.02$ & $\Delta P/L$ & $d > 0.07$m and $\theta > 120^\circ$ & 3696:400\\
        \midrule
        Expansion joint & $0.0005 - 0.005$ & $15 - 60$ & $0.002 - 0.5$ & $f$ & $d > 0.0025$m and $\theta > 50^\circ$ & 1764:589\\
         \bottomrule
       \end{tabular}}
    \label{tab:train_data}
\end{table}

For both the systems described preciously, we have 3-dimensional (3D) features, meaning we vary are 3 input parameters for the CFD simulations. For the bent pipe system, these features are pipe diameter ($d$), bend angle ($\theta$), and average inlet velocity ($\langle v_{in} \rangle$). For the expansion joint, features are inlet pipe diameter ($d$), expansion angle ($\theta$) and average inlet velocity ($\langle v_{in} \rangle$). The target property $y$ for bent pipe is pressure drop $\Delta P/L$ and for the expansion joint, it is the backflow volume percentage  $f$. To start with, a range of acceptable values for the input parameters is chosen to ensure laminar fluid flow and the 3D feature space is divided into training data and testing data. The ranges for features and test/train split criteria are shown in Table~\ref{tab:train_data}. Since we are not tuning hyperparameters, we do not use a validation split. To make sure that physics of the system are obeyed, we define asymptotes based on prior system knowledge.
In the bent pipe system, $d=0$ will result infinite pressure drop ($\displaystyle{\lim_{d\to 0} \frac{\Delta P}{L}(d, \theta, v)= \infty}$), which is a valid asymptote as per the HP equation (Equation~\ref{eq:HP}).
Parameter $\theta$ is set to be bounded between $15-60 ^{\circ}$ in the expansion joint system. The rationale behind this choice comes from the fact that no backflow is formed at smaller expansion angles, which results a zero variance in the labels for all possible variations of the other two features ($f(d, \theta,v) \Big|_{\theta<15^{\circ}} = 0;  \forall d, v$). On the other hand, at larger angles beyond this limit, the size of the system increases significantly as a result of geometrical constraint set for the length of the expansion section (see Figure~\ref{fig:expansion_joint}). Specifically, for the case of $\theta = 90^{\circ}$, the system size becomes infinite as a result of having an infinite outlet diameter.

For AL, three random points $\vec{x}$ from the training data are sampled, and CFD simulations are generated to find corresponding labels $y$. This is our initial training data ($\mathcal{L}$) for our pool-based AL and the rest of the feature points form the unlabeled data pool $\mathcal{U}$. CFD simulations are used to label data for feature points obtained from Equation~\ref{eq:uncertainty_sampling}. After $N$ such queries, feature-label pairs in $\mathcal{L}$ are used as training data for the SISSO algorithm. We then add asymptotic points to this training data. The number of asymptotic points added depends on the number of training data points $N$ in $\mathcal{L}$. We add greater of 3 or 10\% of $N$ points to $\mathcal{L}$ and make sure that all asymptotic conditions are represented. To create aymptotic data points, the primary features apart from the variable for which the aymptote is defined are sampled randomly from the regime defined for that feature. So, for the bent pipe system, when $\displaystyle{\lim_{d\to 0} \frac{\Delta P}{L}(d, \theta, v) = \infty}$, $\theta$ and $\langle v_{in} \rangle$ are randomly sampled from the bounds defined in Table~\ref{tab:train_data}. Density and Viscosity are also added as features for SISSO. We use the operator set $\{ +, -, \times, \div, \exp, -\exp, ()^{-1}, ()^2, \sin, \cos \}$ with our features for both systems and set the number of descriptors to 3. The symbolic equation obtained from SISSO is used to predict labels for the test features.  This method is compared against random search and grid search experiment design algorithms. Random search is random selection of feature points with uniform sampling probability from the data pool. Grid search is when a hypercube of points from the 3D feature grid are selected. Grid is equivalent to a factorial design if we view our levels as discretization of our features. SISSO is used to find equations for both these methods so they can be compared against our AL + SISSO method. The difference in these methods is reported using significance statistics obtained from an independent samples t-test. The independent samples t-test  is a parametric test that compares the means of two independent distributions and gives statistics to confirm the hypothesis that the two populations are significantly different.\cite{Kalpic2011} 

\section*{Results and Discussion}
The method described above, AL + SISSO, is tested and compared with baseline methods used for experiment design like random search and grid search. The objective was to obtain a symbolic relationship between inputs/features and outputs/labels, given data. 

\begin{figure}[!ht]
    \centering
    \includegraphics[width=\textwidth]{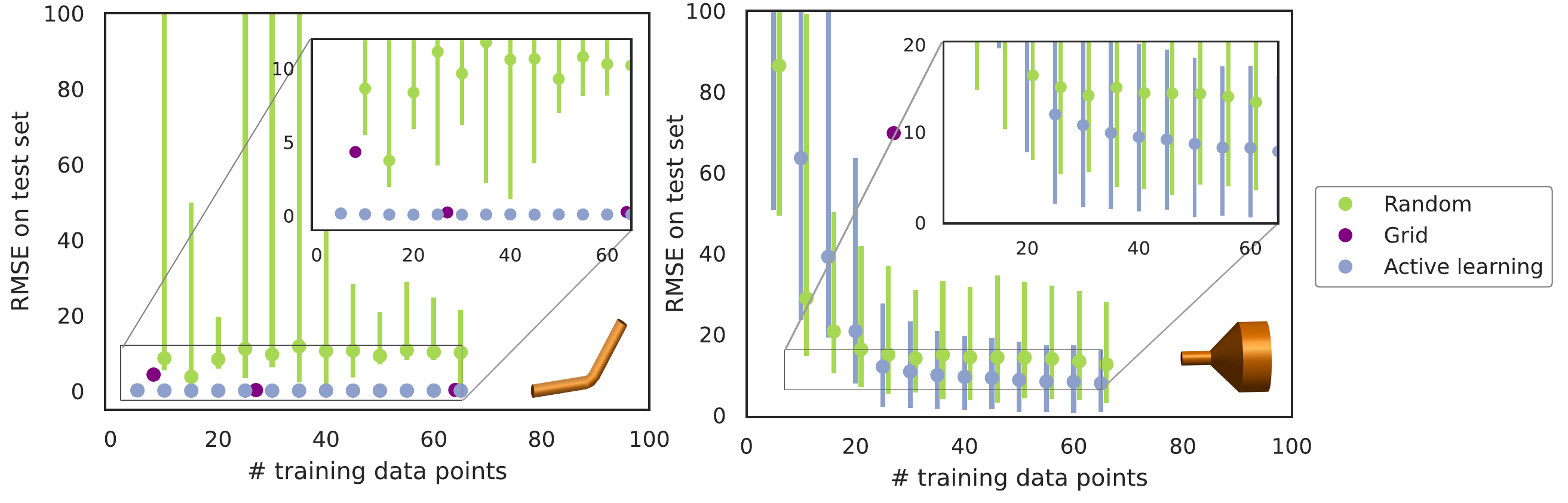}
    \caption{Root mean squared errors (RMSE) as a function of number of training data points. The symbolic regression equations obtained from different models for varying number of training data points are evaluated for test data and the RMSE in predictions is plotted. Error bars on these points indicate the 50$^{th}$ quantile for RMSEs on test data. It is observed that AL model for experiment design has the best performance followed by random selection of experiment points. A grid search model requires many data points to achieve comparable performance.}
    \label{fig:rmse}
\end{figure}

\begin{figure}
    \centering
    \includegraphics[width=\textwidth]{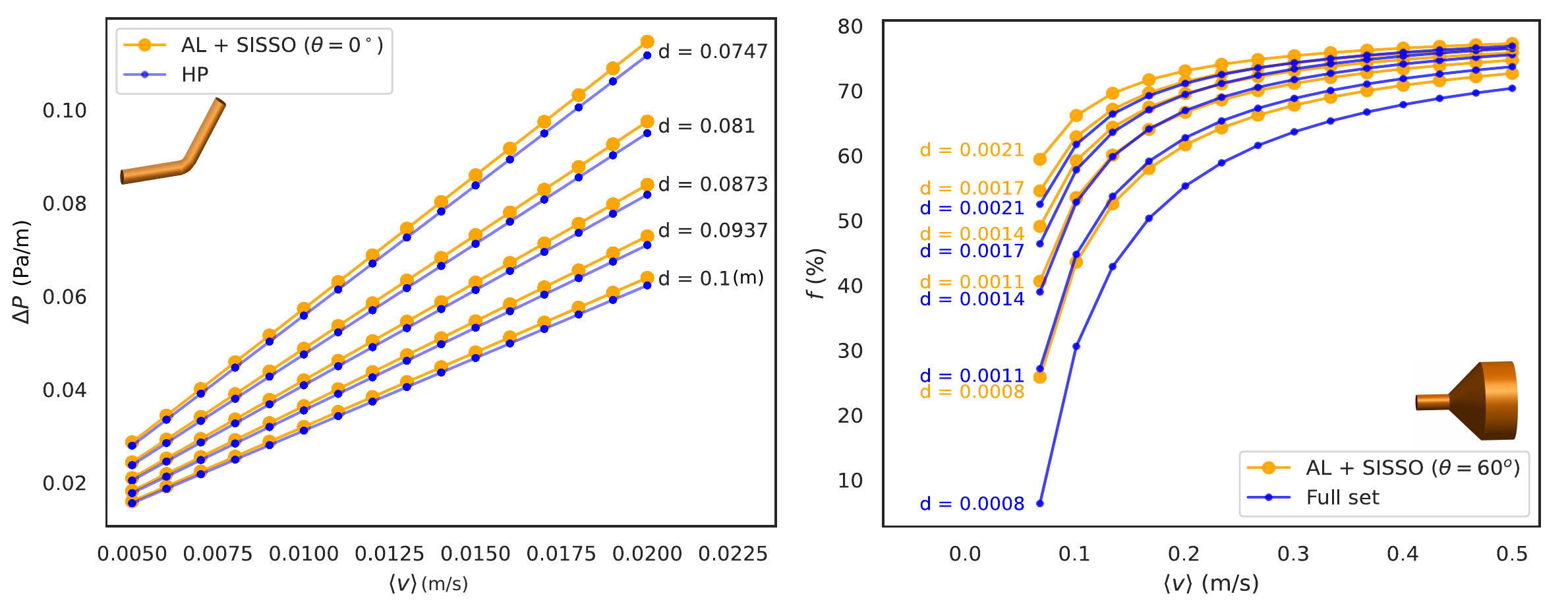}
    \caption{Comparison of SISSO results with baseline models. }
    \label{fig:baseline}
\end{figure}

In Figure~\ref{fig:rmse}, we compare the root mean squared error (RMSE) on test data from SR equations for AL, random search and grid search. The symbolic relationships obtained from SISSO for AL, random search and grid search are evaluated on the test data points and respective RMSEs are calculated between these values and actual CFD values. Test data was withheld from the training data pool for both systems, as shown in Table~\ref{tab:train_data}. In Figure~\ref{fig:rmse}, each data point for AL and random search represents the mean value for RMSE from 100 independent iterations of training data sampling followed by SISSO. The uncertainty in AL comes from the randomly sampled initial training points and is in the coefficients $C_i$ of SISSO equations. For bent pipe, we observe that AL converges quickly and requires fewer training points to get an accurate symbolic equation between features and labels. RMSEs for random search and AL are significantly different ($p = 8.8\times 10^{-23}$, via independent samples t-test). Grid search outperforms random search and requires 27 training points to converge. Table~\ref{tab:system1} shows some equations obtained from SISSO for AL, random search, grid search and the full training set. The complete list of equations obtained for different methods, as a function of training data points, is reported in Table~\ref{tab:system1_si}. The equations reported are those that are observed maximum times (mode of the distribution) for a given value of training points. AL combined with SISSO provides an accurate equation with as low as 10 training data points, and the general form of the equation remains the same with increasing training data points. Random search requires 15 training points to obtain a similar equation. The equation obtained for grid search varies with increasing training points. Although, random search obtains the correct general equation with few points, the variance in the coefficients for these equations is high and hence, have a high RMSE in their approximation of pressure drop in the system.
For the expansion joint, AL outperforms random search and grid search. The difference in RMSE between AL and random search increases as the number of training points increases ($p = 0.0125$, via independent samples t-test). Equations obtained from SISSO for this system are reported in Table~\ref{tab:system2} and the complete list as a function of training data points can be found in Table~\ref{tab:system2_si}. The general equation for AL, random search and grid search remains the same after 30, 60 and 512 training points, respectively.

We also compare the performance of AL + SISSO models with baseline models. Figure~\ref{fig:baseline} shows how AL + SISSO compares to the respective baseline models. At $\theta = 0^\circ$, the bent pipe becomes a straight pipe, for which $\Delta P/L$ can be calculated using the HP equation (Equation~\ref{eq:HP}) and is considered the baseline. AL + SISSO equation results fit the HP equation. For backflow in the expansion joint, there is no such theory-derived equation to compare against. So, we perform SISSO on the entire feature pool and consider that as the baseline model. AL + SISSO predictions for $\theta=60^\circ$ underestimate the backflow percentage compared to the baseline. However, the curves for AL + SISSO follow the same form as the baseline and there is an offset along the y-axis. SISSO equations obtained for AL and the baseline are the same, and the difference in predicted labels comes from the coefficients C1, C2, C3 for the two equations. This is reasonable since our goal is to find a symbolic equation to understand the system and not to minimize the regression error. 

In the final analysis, it is important to consider how well the results of AL+SISSO can be used to describe flow and how well they compare to known equations when they exist. In the case of the bent pipe, the data in Figure 5 confirms that the AL+SISSO matches nearly exactly with the HP Equation. 
Note that all symbolic equations beyond 10 training points shown in Table~\ref{tab:system1_si} are consistent with theory-driven HP equation (Equation~\ref{eq:HP}) for $\theta=0$ and accurately describe the pressure drop in the system. When known equations do not exist, which is the case for most complex flow scenarios, the ability of AL+SISSO to describe flow needs to be carefully interpreted and compared to best known approximations. For the expansion joint, there are no accepted equations for the backflow volume percentage to compare against for any geometry, so comparisons are made with the entire feature pool referred to as the full set in the figure. The data in Figure 5 does show a difference between the AL+SISSO and the full set. The important observation is that the shape of the graph for the volume percentage versus velocity are quite similar, rising and then leveling off with velocity as one would expect. The observed difference is the result of the coefficients in the equations as mentioned above and not an incorrect symbolic equation.

\section*{Conclusions}
We introduce an AL approach combined with SR for obtaining an empirical symbolic relationship between system variables for CFD simulations. This framework eliminates the need for the conventional trial-and-error or grid search methods for picking feature points since we let AL pick these points based on prior information available. We demonstrate the use of this method for two CFD systems and compare them against conventional methods.  The results obtained from SISSO are more interpretable than those obtained from black box functions, and can be directly used. This method also greatly reduces the amount of data needed to get meaningful insights about a CFD system. One limitation of this method is that the obtained symbolic relationships are only valid for fluid flow regimes described by the feature domain considered for the training data pool (i.e. laminar flow regime for the two examples illustrated). Augmenting training data with asymptotic points from prior scientific knowledge helps ensure that the equations obey the physics of the system.

\renewcommand*{\arraystretch}{1.2}
\begin{longtable}{p{2cm}p{2cm}p{2cm}p{8cm}}
\caption{Equations obtained from SISSO for the Bent Pipe system} \label{tab:system1} \\
\toprule
Method &  Training Points & Mean Test RMSE & Equation \\
\midrule
\endfirsthead
\toprule
Method &  Training Points & Mean Test RMSE & Equation \\
\midrule
\endhead
\midrule
\multicolumn{3}{r}{{Continued on next page}} \\
\midrule
\endfoot
\bottomrule
\endlastfoot
    AL & 5 & 0.143 & $ \frac{C_{1} v + d \left(C_{2} v \sin{\left(\theta \right)} + C_{3} \theta \cos{\left(\theta \right)}\right)}{d^{2}}$ \\ 
Random & 5 & 349.42 & $ \frac{C_{1} \cos{\left(d \right)}}{d^{2}} + \frac{C_{2} e^{- v}}{d^{2}} + \frac{C_{3} \theta v^{2}}{d}$ \\ 
Grid & 8 & 4.32 & $ \frac{C_{1} v}{d^{2}} + \frac{C_{2} \theta^{2} v}{d} - C_{3} e^{v} + C_{3} e^{- \theta}$ \\
AL & 60 & 0.072 & $ \frac{v \left(C_{1} + C_{2} \theta + C_{3} d \theta v\right)}{d^{2}}$ \\ 
Random & 60 & 10.29 & $ \frac{v \left(C_{1} + C_{2} \theta + C_{3} d \theta v\right)}{d^{2}}$ \\ 
Grid & 64 & 0.246 & $ \frac{C_{1}   v + C_{2} d \theta v^{2} + C_{3} \theta^{2}}{d^{2}}$ \\ 
Full set & 3696 & 0.085 & $ \frac{v \left(C_{1} + C_{2} \theta + C_{3} d \theta v\right)}{d^{2}}$ \\ 

\end{longtable}

\renewcommand*{\arraystretch}{1.2}
\begin{longtable}{p{2cm}p{2cm}p{2cm}p{8cm}} 
\caption{Equations obtained from SISSO for the Expansion Joint} \label{tab:system2}\\
\toprule
Method &  Training Points  & Mean Test RMSE &Equation \\
\midrule
\endfirsthead
\toprule
Method &  Training Points & Mean Test RMSE & Equation \\
\midrule
\endhead
\midrule
\multicolumn{3}{r}{{Continued on next page}} \\
\midrule
\endfoot
\bottomrule
\endlastfoot
    Random & 5 & 86.42 & $ C_{1} v \sin{\left(\theta \right)} + C_{2} \theta v e^{d} + C_{3} d e^{d}$ \\ 
AL & 5 & 107.13 & $ C_{1} \sin{\left(\theta \right)} + \frac{C_{2} \theta}{v \sin{\left(d \right)}} + C_{3} \sin{\left(\frac{\theta}{d} \right)}$ \\
Grid & 8 & 4955.92 & $ \theta \left(C_{1} v \sin{\left(\theta \right)} + C_{2} d \theta v + \frac{C_{3} d}{\cos{\left(d \right)}}\right)$ \\
AL & 60 & 8.25 & $ \frac{\theta \left(C_{1} d v \cos{\left(\theta \right)} + C_{2} + C_{3} \theta\right)}{d v}$ \\ 
Random & 60 & 13.30 & $ C_{1} e^{- d \theta} + C_{2} e^{\theta} + C_{2} e^{- \theta} + C_{3} d v e^{- d}$ \\ 
Grid & 64 & 79.35 & $ C_{1} d \theta e^{- d} + C_{2} d^{3} v + C_{3} \sin{\left(  \theta \right)}$ \\
Full set & 1764 & 12.65 & $ \frac{C_{1} \theta d v \cos{\left(\theta \right)} + C_{2} d v \sin{\left(d \theta \right)} + C_{3} }{d v}$ \\ 

\end{longtable}

\section*{Acknowledgements}

We thank the Center for Integrated Research Computing (CIRC) at University of Rochester for providing computational resources and technical support. This material is based upon work supported by the National Science Foundation (NSF) under Grant 1751471, the Molecular Sciences Software Institute (MolSSI) under NSF grant OAC-1547580, and the Maximizing Investigators’ Research Award Grant R35 GM137966 by the National Institute of General Medical Sciences under the National Institutes of Health.

\bibliography{bibliography}
\bibliographystyle{unsrtnat}

\newpage
\begin{center}
\LARGE \textsc{ Supplemental Information: Iterative Symbolic Regression for Learning Transport Equations}
\end{center}
\setcounter{section}{0}
\setcounter{equation}{0}
\setcounter{figure}{0}
\setcounter{table}{0}
\setcounter{page}{1}
\makeatletter
\renewcommand{\thesection}{S\arabic{section}}
\renewcommand{\theequation}{S\arabic{equation}}
\renewcommand{\thefigure}{S\arabic{figure}}
\renewcommand{\bibnumfmt}[1]{[S#1]}
\renewcommand{\citenumfont}[1]{S#1}

\section{CFD Models Methods}
\subsection{Bent Pipe System} \label{section1.1}
By cutting the geometry in half, a symmetrical boundary condition is imposed on the front interior face. To avoid invalid geometries, a set of parametric constraints are applied, where both inlet and outlet side extrusions have length of 5$d$. The reference point for the revolving bend is also set to be at a distance $d$ from the pipe center-line. Given the nature of our automated workflow, the mesh sizing is also constrained to $d/20$ which enables adjustable meshing based on the user input for the pipe diameter.

\subsection{Expansion Joint System} \label{section1.2}
The inlet-side extrusion length, expansion height and outlet-side extrusion length are set to be 2$d$, while the expansion angle is $\theta$.
The backflow volume is calculated by summing over the volume of mesh elements where $v_z < 0$:
\begin{equation}
\label{eq:back_flow_v}
    V_{bf}= \int_{V} \phi_V dV
\end{equation}
where $\phi$ is our user-defined function (UDF):
\begin{equation}
    \label{eq:phi}
        \phi_V = 
    \begin{cases}
        1& \text{if } v_z< -0.0001 \\
        0  & \text{otherwise}
    \end{cases}
\end{equation}
Note that in Equation~\ref{eq:phi}, the threshold is set to a small negative value (-0.0001 m/s), to avoid the inclusion of those mesh elements at the wall, where no-slip boundary is applied.

\section{Active learning Softmax Adjustment} \label{sectionAL}

\begin{figure} [!h]
    \centering
    \includegraphics[width=0.75\textwidth]{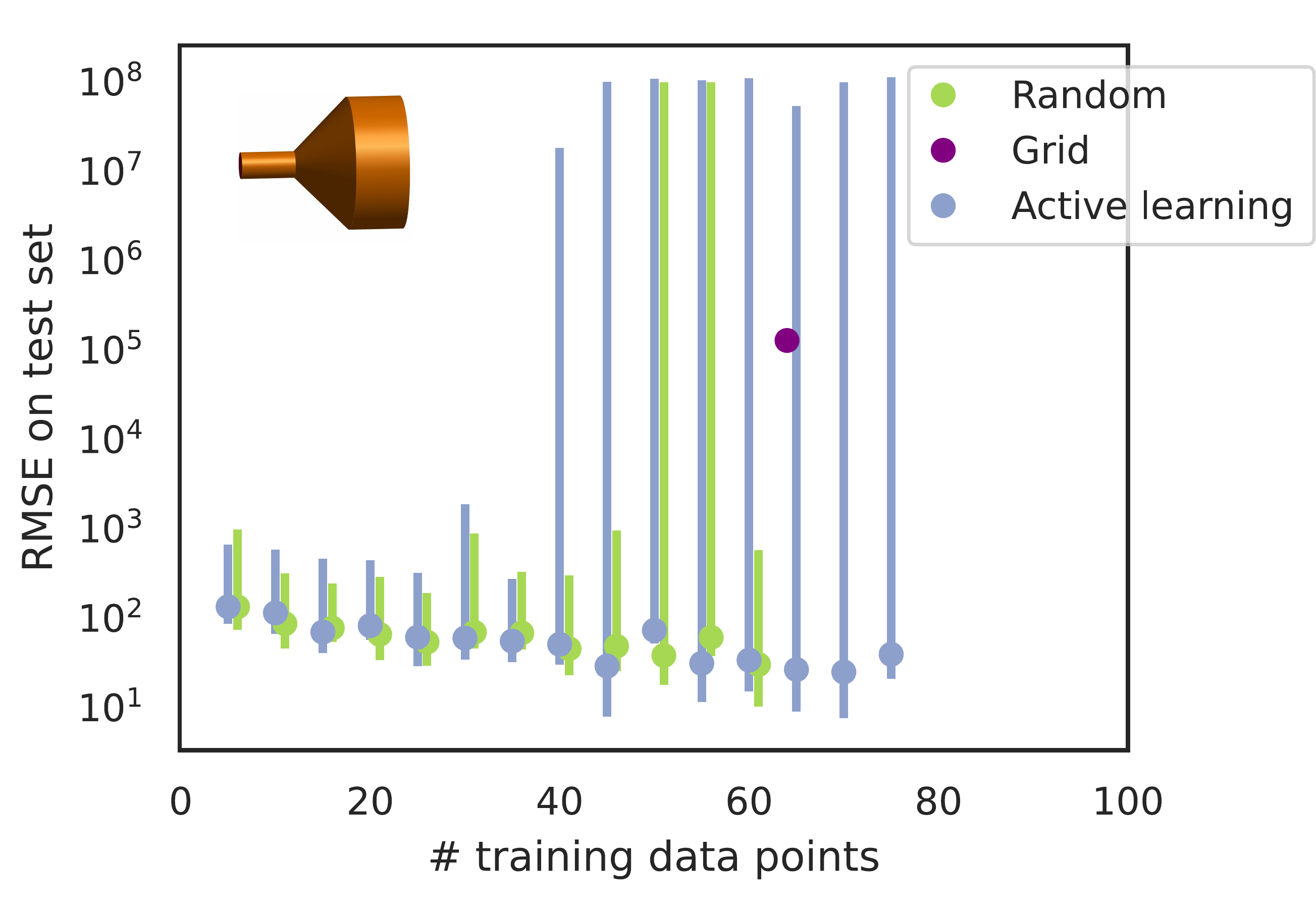}
    \caption{RMSE on test data for when AL is used without the softmax trick.}
    \label{fig:without_softmax}
\end{figure}

When uncertainty prediction for AL uses argmax (Equation~\ref{eq:wo_softmax}), the predictions are sensitive to noise in the labels. Figure~\ref{fig:without_softmax} shows the RMSE on test data when softmax is not applied to AL prediction uncertainty.

\begin{equation} 
    \vec{x}_{i+1} = \argmax_x\ \left[1 - max\left(P_{\Phi}(\hat{y}|\vec{x})\right)\right]
    \label{eq:wo_softmax} 
\end{equation}

\section{Symbolic Regression Equations}

\renewcommand*{\arraystretch}{1.5}
\begin{longtable}{p{2cm}p{2cm}p{2cm}p{8cm}}
\caption{Equations obtained from SISSO for the Bent Pipe system} \label{tab:system1_si} \\
\toprule
Method &  Training Points & Mean Test RMSE & Equation \\
\midrule
\endfirsthead
\toprule
Method &  Training Points & Mean Test RMSE & Equation \\
\midrule
\endhead
\midrule
\multicolumn{3}{r}{{Continued on next page}} \\
\midrule
\endfoot
\bottomrule
\endlastfoot
    AL & 5 & 0.143 & $ \frac{C_{1} v + d \left(C_{2} v \sin{\left(\theta \right)} + C_{3} \theta \cos{\left(\theta \right)}\right)}{d^{2}}$ \\ 
Random & 5 & 349.42 & $ \frac{C_{1} \cos{\left(d \right)}}{d^{2}} + \frac{C_{2} e^{- v}}{d^{2}} + \frac{C_{3} \theta v^{2}}{d}$ \\ 
Grid & 8 & 4.32 & $ \frac{C_{1} v}{d^{2}} + \frac{C_{2} \theta^{2} v}{d} - C_{3} e^{v} + C_{3} e^{- \theta}$ \\ 
Random & 10 & 8.63 & $ C_{1} v^{3} + C_{2} v^{3} + \frac{C_{3} \theta v}{d^{2}}$ \\
AL & 10 & 0.095& $ \frac{v \left(\frac{C_{1} d}{\sin{\left(d \right)}} + C_{2} \theta + C_{3} d \theta v\right)}{d^{2}}$ \\ 
Random & 15 & 3.73 & $ \frac{v \left(C_{1} + C_{2} \theta + C_{3} d \theta v\right)}{d^{2}}$ \\ 
AL & 15 & 0.068 & $ \frac{v \left(\frac{C_{1}}{\sin{\left(d \right)}} + C_{2} \theta v + C_{3} d \theta^{2} v\right)}{d}$ \\ 
AL & 20 & 0.065 & $ \frac{C_{1} \sin{\left(v \right)}}{d^{2}} + \frac{C_{2} \theta v^{2}}{d} + C_{3} v^{2} e^{\theta}$ \\ 
Random & 20 & 8.36 & $ \frac{v \left(C_{1}   + C_{2} \theta + C_{3} d \theta v\right)}{d^{2}}$ \\ 
AL & 25 & 0.075 & $ \frac{C_{1} \sin{\left(v \right)}}{d^{2}} + \frac{C_{2} \theta v^{2}}{d} + C_{3} v^{2} e^{\theta}$ \\ 
Random & 25 & 11.13 & $ \frac{v \left(C_{1} + C_{2} \theta + C_{3} d \theta v\right)}{d^{2}}$ \\ 
Grid & 27 & 0.22 & $ \frac{C_{1} v + C_{2} \theta^{2} + C_{3} d \theta v^{2}}{d^{2}}$ \\ 
Random & 30 & 9.65 & $ \frac{v \left(C_{1} + C_{2} \theta + C_{3} d \theta v\right)}{d^{2}}$ \\ 
AL & 30 & 0.061 & $ \frac{v \left(\frac{C_{1} d}{\sin{\left(d \right)}} + C_{2} \theta + C_{3} d \theta v\right)}{d^{2}}$ \\ 
AL & 35 & 0.071 & $ \frac{v \left(C_{1}   + C_{2} \theta + C_{3} d \theta v\right)}{d^{2}}$ \\ 
Random & 35 & 11.77 & $ \frac{C_{1} \sin{\left(v \right)} + C_{2} \theta v + C_{3} d \theta v^{2}}{d^{2}}$ \\ 
Random & 40 & 10.59 & $ \frac{v \left(C_{1} + C_{2} \theta + C_{3} d \theta v\right)}{d^{2}}$ \\ 
AL & 40 & 0.083 & $ \frac{v \left(\frac{C_{1} d}{\sin{\left(d \right)}} + C_{2} \theta + C_{3} d \theta v\right)}{d^{2}}$ \\ 
Random & 45 & 10.65 & $ \frac{v \left(C_{1} + C_{2} \theta + C_{3} d \theta v\right)}{d^{2}}$ \\ 
AL & 45 & 0.066 & $ \frac{C_{1} \sin{\left(v \right)}}{d^{2}} + \frac{C_{2} \theta v^{2}}{d} + C_{3} v^{2} e^{\theta}$ \\ 
AL & 50 & 0.0822 & $ \frac{v \left(\frac{C_{1} d}{\sin{\left(d \right)}} + C_{2} \theta + C_{3} d \theta v\right)}{d^{2}}$ \\ 
Random & 50 & 9.28 & $ \frac{v \left(C_{1} + C_{2} \theta + C_{3} d \theta v\right)}{d^{2}}$ \\ 
Random & 55 & 10.78 & $ \frac{v \left(C_{1} + C_{2} \theta + C_{3} d \theta v\right)}{d^{2}}$ \\ 
AL & 55 & 0.075 & $ \frac{v \left(\frac{C_{1} d}{\sin{\left(d \right)}} + C_{2} \theta + C_{3} d \theta v\right)}{d^{2}}$ \\ 
Random & 60 & 10.29 & $ \frac{v \left(C_{1} + C_{2} \theta + C_{3} d \theta v\right)}{d^{2}}$ \\ 
AL & 60 & 0.072 & $ \frac{v \left(C_{1} + C_{2} \theta + C_{3} d \theta v\right)}{d^{2}}$ \\ 
Grid & 64 & 0.246 & $ \frac{C_{1}   v + C_{2} d \theta v^{2} + C_{3} \theta^{2}}{d^{2}}$ \\ 
AL & 65 & 0.076 & $ \frac{v \left(C_{1} + C_{2} \theta + C_{3} d \theta v\right)}{d^{2}}$ \\ 
Random & 65 & 10.21 & $ \frac{v \left(\frac{C_{1} d}{\sin{\left(d \right)}} + C_{2} \theta + C_{3} d \theta v\right)}{d^{2}}$ \\ 
Grid & 125 & 2.92 & $ \frac{v \left(C_{1}   + C_{2} d \theta v + C_{3} d^{3} \theta v\right)}{d^{2}}$ \\ 
Random & 125 & 9.91 & $ \frac{v \left(C_{1} + C_{2} \theta + C_{3} d \theta v\right)}{d^{2}}$ \\ 
Grid & 216 & 1.07 & $ \frac{v \left(C_{1}   + C_{2} d \theta v + C_{3} d^{2} \theta^{2}\right)}{d^{2}}$ \\ 
Random & 216 & 9.97 & $ \frac{v \left(C_{1} + C_{2} \theta + C_{3} d \theta v\right)}{d^{2}}$ \\ 
Grid & 343 & 0.38 & $ \frac{C_{1}   v}{d^{2}} + \frac{C_{2} v \sin{\left(\theta \right)}}{d} + \frac{C_{3} \theta e^{- \theta}}{v}$ \\ 
Random & 343 & 10.12 & $ \frac{v \left(C_{1} + C_{2} \theta + C_{3} d \theta v\right)}{d^{2}}$ \\ 
Grid & 512 & 0.10 & $ \frac{C_{1} v}{d^{2}} + \frac{C_{2} \theta e^{- \theta}}{d} + \frac{C_{3} v \sin{\left(\theta \right)}}{d}$ \\ 
Random & 512 & 10.36 & $ \frac{v \left(C_{1} + C_{2} \theta + C_{3} d \theta v\right)}{d^{2}}$ \\ 
Random & 729 & 10.64 & $ \frac{v \left(C_{1} + C_{2} \theta + C_{3} d \theta v\right)}{d^{2}}$ \\ 
Grid & 729 & 0.10 & $ \frac{C_{1} \sin{\left(v \right)}}{d^{2}} + \frac{C_{2} \theta e^{- \theta}}{d} + \frac{C_{3} v \sin{\left(\theta \right)}}{d}$ \\ 
Full set & 3696 & 0.085 & $ \frac{v \left(C_{1} + C_{2} \theta + C_{3} d \theta v\right)}{d^{2}}$ \\ 

\end{longtable}

\renewcommand*{\arraystretch}{1.5}
\begin{longtable}{p{2cm}p{2cm}p{2cm}p{8cm}} 
\caption{Equations obtained from SISSO for the Expansion Joint} \label{tab:system2_si}\\
\toprule
Method &  Training Points  & Mean Test RMSE &Equation \\
\midrule
\endfirsthead
\toprule
Method &  Training Points & Mean Test RMSE & Equation \\
\midrule
\endhead
\midrule
\multicolumn{3}{r}{{Continued on next page}} \\
\midrule
\endfoot
\bottomrule
\endlastfoot
    Random & 5 & 86.42 & $ C_{1} v \sin{\left(\theta \right)} + C_{2} \theta v e^{d} + C_{3} d e^{d}$ \\ 
AL & 5 & 107.13 & $ C_{1} \sin{\left(\theta \right)} + \frac{C_{2} \theta}{v \sin{\left(d \right)}} + C_{3} \sin{\left(\frac{\theta}{d} \right)}$ \\ 
Grid & 8 & 4955.92 & $ \theta \left(C_{1} v \sin{\left(\theta \right)} + C_{2} d \theta v + \frac{C_{3} d}{\cos{\left(d \right)}}\right)$ \\ 
Random & 10 & 28.86 & $ C_{1} \sin{\left( v \right)} + C_{2} d \sin{\left(\theta \right)} + C_{3} d \theta e^{- d}$ \\ 
AL & 10 & 63.54 & $ C_{1} \theta^{2} e^{\theta} + C_{2} \theta^{3} + \frac{C_{3} \sin{\left(\theta \right)}}{d v}$ \\ 
Random & 15 & 20.68 & $ C_{1} d \theta e^{\theta} + C_{2} \sin{\left(d \theta \right)} + C_{3} \theta v e^{- d}$ \\ 
AL & 15 & 39.14 & $ \frac{\frac{C_{1} d v \sin{\left(2 \theta \right)}}{2} + C_{2} \theta + C_{3} \theta^{2}}{d v}$ \\ 
Random & 20 & 16.26 & $ C_{1} \sin{\left( v \right)} + C_{2} d \theta \cos{\left(\theta \right)} + C_{3} d \theta e^{- d}$ \\ 
AL & 20 & 20.75 & $ \frac{\frac{C_{1} d v \sin{\left(2 \theta \right)}}{2} + C_{2} \theta + C_{3} \theta^{2}}{d v}$ \\ 
Random & 25 & 14.95 & $ C_{1} \sin{\left( v \right)} + C_{2} d \theta \cos{\left(\theta \right)} + C_{3} d \theta e^{- d}$ \\ 
AL & 25 & 11.94 & $ \frac{\frac{C_{1} d v \sin{\left(2 \theta \right)}}{2} + C_{2} \theta + C_{3} \theta^{2}}{d v}$ \\ 
Grid & 27 & 69.72 & $ \left(C_{1} d \theta + v \left(C_{2} \theta + C_{3}\right) e^{\theta} \cos{\left(d \right)}\right) e^{- \theta}$ \\ 
Random & 30 & 14.03 & $ C_{1} \sin{\left( v \right)} + C_{2} d \theta \cos{\left(\theta \right)} + C_{3} d \theta e^{- d}$ \\ 
AL & 30 & 10.75 & $ \frac{\theta \left(C_{1} d v \cos{\left(\theta \right)} + C_{2} + C_{3} \theta\right)}{d v}$ \\ 
AL & 35 & 9.91 & $ \frac{\theta \left(C_{1} d v \cos{\left(\theta \right)} + C_{2} + C_{3} \theta\right)}{d v}$ \\ 
Random & 35 & 14.91 & $ C_{1} d \theta e^{\theta} + C_{2} \sin{\left(d \theta \right)} + C_{3} v e^{- d}$ \\ 
AL & 40 & 9.44 & $ \frac{\theta \left(C_{1} d v \cos{\left(\theta \right)} + C_{2} + C_{3} \theta\right)}{d v}$ \\ 
Random & 40 & 14.31 & $ \theta \left(C_{2} v + d \left(C_{1} \cos{\left(\theta \right)} + C_{3} \sin{\left(d \right)}\right) e^{d}\right) e^{- d}$ \\ 
AL & 45 & 9.18 & $ \frac{\theta \left(C_{1} d v \cos{\left(\theta \right)} + C_{2} + C_{3} \theta\right)}{d v}$ \\ 
Random & 45 & 14.26 & $ \frac{\frac{C_{1} d v \sin{\left(2 \theta \right)}}{2} + C_{2} \theta^{2} + C_{3} \theta}{d v}$ \\ 
AL & 50 & 8.69 & $ \frac{\theta \left(C_{1} d v \cos{\left(\theta \right)} + C_{2} + C_{3} \theta\right)}{d v}$ \\ 
Random & 50 & 14.22 & $ \theta \left(C_{2} v + d \left(C_{1} \cos{\left(\theta \right)} + C_{3} \sin{\left(d \right)}\right) e^{d}\right) e^{- d}$ \\ 
AL & 55 & 8.28 & $ \frac{\theta \left(C_{1} d v \cos{\left(\theta \right)} + C_{2} + C_{3} \theta\right)}{d v}$ \\ 
Random & 55 & 13.91& $ C_{1} e^{- d \theta} + C_{2} e^{\theta} + C_{2} e^{- \theta} + C_{3} d v e^{- d}$ \\ 
Random & 60 & 13.30 & $ C_{1} e^{- d \theta} + C_{2} e^{\theta} + C_{2} e^{- \theta} + C_{3} d v e^{- d}$ \\ 
AL & 60 & 8.25 & $ \frac{\theta \left(C_{1} d v \cos{\left(\theta \right)} + C_{2} + C_{3} \theta\right)}{d v}$ \\ 
Grid & 64 & 79.35 & $ C_{1} d \theta e^{- d} + C_{2} d^{3} v + C_{3} \sin{\left(  \theta \right)}$ \\ 
Random & 65 & 12.55 & $ C_{1} e^{- d \theta} + C_{2} e^{\theta} + C_{2} e^{- \theta} + C_{3} d v e^{- d}$ \\ 
AL & 65 & 7.86 & $ \frac{\theta \left(C_{1} d v \cos{\left(\theta \right)} + C_{2} + C_{3} \theta\right)}{d v}$ \\ 
Grid & 125 & 67.13& $ C_{1} \theta v \cos{\left(\theta \right)} + C_{2} \theta v^{2} + C_{3} v e^{- d}$ \\ 
Grid & 216 & 126.55 & $ \theta \left(C_{3} v + d \left(C_{1} \cos{\left(\theta \right)} + C_{2} d\right) e^{d}\right) e^{- d}$ \\ 
Grid & 343 & 79.89 & $ C_{1} d \theta \cos{\left(\theta \right)} + C_{2} d^{2} \sin{\left(\theta \right)} + C_{3} \theta v e^{- d}$ \\ 
Grid & 512 & 8.65 & $ C_{1} d \theta e^{- \theta} + C_{2} d \theta e^{- d} + C_{3} v \cos{\left(\theta \right)}$ \\ 
Grid & 729 & 7.99 & $ C_{1} d \theta e^{- \theta} + C_{2} d \theta e^{- d} + C_{3} v \cos{\left(\theta \right)}$ \\ 
Full set & 1764 & 12.65 & $ C_{1} \theta \cos{\left(\theta \right)} + C_{2} \sin{\left(d \theta \right)} + \frac{C_{3} }{d v}$ \\ 

\end{longtable}

\end{document}